\documentclass[twocolumn,aps,prb,groupedaddress,showpacs,floatfix]{revtex4}
\usepackage{amsmath,amssymb}
\usepackage{graphicx}
\usepackage{dcolumn}
\usepackage{enumerate}
\usepackage{hyperref}

\newcommand{\ZnMnO}{{$\mathrm{Zn}_{1{-}x}\mathrm{Mn}_x\mathrm{O}$}}

\newcommand{\dms}[3]{{$\mathrm{#1}_{1{-}x}\mathrm{#2}_x\mathrm{#3}$}}
\newcommand{\kB}{k_\mathrm{B}}
\newcommand{\muB}{\mu_\mathrm{B}}
\newcommand{\kOe}{\;\mathrm{kOe}}
\newcommand{\K}{\;\mathrm{K}}
\newcommand{\mK}{\;\mathrm{mK}}
\newcommand{\J}[1]{{$J^{(#1)}$}}
\begin{document}
\bibliographystyle{apsrev}
\title{Magnetization steps in $\mathbf{Zn_{1{-}x}Mn_xO}$:
Four largest exchange constants and single-ion anisotropy}
\author{X. Gratens}
\author{V. Bindilatti}
\email{vbindilatti@if.usp.br}
\author{N. F. Oliveira, Jr.}
\affiliation{Instituto de F{\'\i}sica, Universidade de S{\~a}o Paulo,\\
 C. Postal 66.318, 05315--970 S{\~a}o Paulo--SP, Brazil}
\author{Y. Shapira}
\email{yshapira@granite.tufts.edu} \affiliation{Department of
Physics and Astronomy, Tufts University, Medford, MA 02155}
\author{S. Foner}
\affiliation{Francis Bitter Magnet Laboratory, Massachusetts
Institute of Technology, Cambridge, MA 02139}
\author{Z. Golacki}
\affiliation{Institute of Physics, Polish Academy of Sciences, Al.
Lotnikow 32/46, Pl. 02--668, Warsaw, Poland}
\author{T. E. Haas}
\affiliation{Department of Chemistry, Tufts University, Medford,
MA 02155}

\date{\today}

\begin{abstract}
Magnetization steps (MST's) from $\mathrm{Mn}^{2+}$ pairs in
several single crystals of
$\mathrm{Zn}_{1{-}x}\mathrm{Mn}_x\mathrm{O}$
($0.0056{\le}x{\le}0.030$), and in one powder ($x{=}0.029$), were
observed. They were used to determine the four largest exchange
constants (largest $J$'s), and the single-ion axial anisotropy
parameter, $D$. The largest two exchange constants,
$J_1/k_\mathrm{B}{=}{-}18.2{\pm}0.5\;\mathrm{K}$ and
$J_1^\prime/k_\mathrm{B}{=}{-}24.3{\pm}0.6\;\mathrm{K}$, were obtained
from large peaks in the differential susceptibility, $dM/dH$,
measured in pulsed magnetic fields, $H$,  up to
$500\;\mathrm{kOe}$. These two largest $J$'s are associated with
the two inequivalent classes of nearest neighbors (NN's) in the
wurtzite structure. The 29\% difference between $J_1$ and
$J_1^\prime$ is substantially larger than 13\% in \dms{Cd}{Mn}{S},
and 15\% in \dms{Cd}{Mn}{Se}. The pulsed-field data also indicate
that, despite the direct contact between the samples and a
superfluid-helium bath, substantial departures from thermal
equilibrium occurred during the $7.4\;\mathrm{ms}$ pulse. The
third- and fourth-largest $J$'s were determined from the
magnetization $M$ at $20\;\mathrm{mK}$, measured in  dc magnetic
fields $H$ up to $90\;\mathrm{kOe}$. Both field orientations
$\mathbf{H}\parallel\mathbf{c}$ and
$\mathbf{H}\parallel[10\bar{1}0]$ were studied. (The
$[10\bar{1}0]$ direction is perpendicular to the $c$-axis,
$[0001]$.) By definition, neighbors which are not NN's are distant
neighbors (DN's). The largest DN exchange constant (third-largest
overall), has the value
$J/k_\mathrm{B}{=}{-}0.543{\pm}0.005\;\mathrm{K}$, and is associated
with the DN at $\mathbf{r}{=}\mathbf{c}$. Because this is not the
closest DN, this result implies that the $J$'s do not decrease
monotonically with the distance $r$. The second-largest DN
exchange constant (fourth-largest overall), has the value
$J/k_\mathrm{B}{\approx}{-}0.080\;\mathrm{K}$. It is  associated
with one of the two  classes of neighbors that have a coordination
number $z_n{=}12$, but the evidence is insufficient for a definite
unique choice. The  dependence of $M$ on the direction of
$\mathbf{H}$ gives $D/k_\mathrm{B} {=}{-}0.039{\pm}0.008\;\mathrm{K}$,
in fair agreement with ${-}0.031\;\mathrm{K}$ from earlier EPR
work.
\end{abstract}

\maketitle
\section{INTRODUCTION}\label{s:intro}
The most extensively studied diluted magnetic semiconductors
(DMS's) are II-VI materials ($\mathrm{A^{II}–B^{VI}}$, where
$\mathrm{A {=} Zn, Cd}$, $\mathrm{B {=} S, Se, Te}$) in which some
of the cations have been replaced by manganese.\cite{note1} The
magnetization-step (MST) method is one of the most effective
techniques of measuring antiferromagnetic (AF) exchange constants
in DMS's.\cite{Shapira02jap,note3} This technique has been used to
determine nearest-neighbor (NN) and distant neighbor (DN) exchange
constants in several II-VI DMS's with the
zinc-blende\cite{Foner89,Bindi98prl,Malarenko00} and
wurtzite\cite{Shapira89,Bindi92,Cajacuri01} structures. Relevant
theoretical treatments of these exchange constants include those
in Refs.~\onlinecite{Larson88,Yu95,Wei93}. In addition to exchange
constants, the MST method gives information about magnetic
anisotropies, and about the distribution of the magnetic ions in
the crystal, on a length scale of several atomic dimensions.

A new class of II-VI DMS's based on ZnO, especially \ZnMnO, has attracted
attention recently because theoretical calculations suggested the possibility
of ferromagnetism above $300\;\mathrm{K}$ in p-type samples.\cite{Dietl00}
Experimental works on epitaxial thin films of \ZnMnO\ gave different results:
ferromagnetism was reported in Ref.~\onlinecite{Jung02}, but according to
Ref.~\onlinecite{Fukumura01} the largest exchange constant is
antiferromagnetic, $J/\kB{\approx}{-}15\;\mathrm{K}$.
In the present work, MST's from several \ZnMnO\ single crystals, and from one
powder sample, were used to determine the largest four exchange constants.
The single-ion axial anisotropy parameter $D$ was also determined.

\section{CRYSTAL STRUCTURE AND CLASSIFICATION OF NEIGHBORS}
\label{s:structure}
\subsection{Classification of Neighbors by Classes}
\label{ss:classification}
\begin{figure}
\includegraphics[width=70mm, keepaspectratio=true, clip]{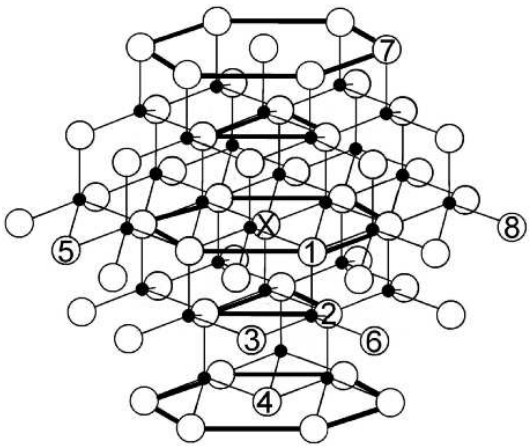}\\
\caption{\label{f:wurtzite}
The wurtzite crystal structure.
The large white spheres are the cations, the small black spheres are
the anions.
The ``central cation'' is labeled as \textsf{X}.
One example of each of the neighbor classes in Table~\ref{table1} is indicated
by the number $n$ specifying that  class.}
\end{figure}

\begin{table}
\caption{\label{table1}
Classification of neighbors in the vicinity of a ``central cation'' in the
hcp cation structure.
The neighbor  class is specified by $n$.
The distance of such a neighbor from the central cation, in the ideal hcp
structure, is $r_n$.
The coordination number $z_n$ is the number of neighbors of  class $n$
which surround the central cation.
The fourth row gives alternative designations for the exchange constants
$J(n)$, e.g., the exchange constant $J(4)$ for a neighbor of the class
$n{=}4$  is designated as $J_3^\prime$.
The superscripts ``in'' and ``out'' distinguish between equidistant but
inequivalent neighbors: those in the same $c$-plane and those in different
$c$-planes.
The dipole-dipole interaction constant $(g\muB)^2/r_n^3$, expressed in kelvin,
is for the lattice parameter $a$ of ZnO, but using the ideal ratio
$c/a {=}\sqrt{8/3}$.
}
\begin{ruledtabular}
\begin{tabular}{c|c|c|c|c|c|c|c|c}
$n$   &   $1$&  $2$&  $3$&  $4$&  $5$&  $6$&  $7$&  $8$\\ \hline
$r_n/a$   &\multicolumn{2}{c|}{$1$}
        &$\sqrt{2}$
        &$\sqrt{{8}/{3}}$
        &\multicolumn{2}{c|}{$\sqrt{3}$}
        &$\sqrt{{11}/{3}}$
        &$2$\\ \hline
$z_n$   &$6$&$6$&$6$&$2$&$6$&$12$&$12$&$6$\\ \hline
$J(n)$   &$J_1^\mathrm{in}$
        &$J_1^\mathrm{out}$
        &$J_2$
        &$J_3^\prime$
        &$J_3^\mathrm{in}$
        &$J_3^\mathrm{out}$
        &$J_4^\prime$
        &$J_4$\\ \hline
$\displaystyle\frac{(g\muB)^2}{r_n^3}$
        &\multicolumn{2}{c|}{$0.073$}
        &$0.026$
        &$0.017$
        &\multicolumn{2}{c|}{$0.014$}
        &$0.010$
        &$0.009$\\
\end{tabular}
\end{ruledtabular}
\end{table}

The hexagonal wurtzite structure of ZnO (space group $P6_3mc$) is shown in
Fig.~\ref{f:wurtzite}.
The cations, (open circles) form an hexagonal close-packed (hcp) structure.
The cation marked by \textsf{X} is chosen as the ``central cation.''
The other cations are often classified by their distances from the central
cation, i.e., nearest neighbors (NN's), second-neighbors, etc.\cite{Larson88}
A major shortcoming of the classification by distance, is that in the (ideal)
hcp structure some equidistant cations are not equivalent from symmetry
point of view.
Equidistant but symmetry-inequivalent cations have different isotropic
exchange constants.

The classification of neighbors by symmetry, instead of  distance, is
discussed in Refs.~\onlinecite{Shapira02jap} and \onlinecite{note3}.
In this classification, neighbors are divided into ``classes.''
Neighbors of the same class have the following property:
When cation sites are occupied by magnetic ions, neighbors of the same
class have the same isotropic (Heisenberg) exchange interaction
with the magnetic ion at the central site.
The exchange constant $J$ is therefore the same for all neighbors of the
same class.
The underlying reason is that all pairs of cation sites consisting of the
central site and a neighbor of a given class are related to each other by
operations of the space group of the cation structure.
Interactions other than isotropic exchange sometimes require distinctions
between neighbors of the same class.\cite{Ladizhansky97}

Properties of several classes of neighbors are given in Table~\ref{table1}.
The number $n$, which is the same as in Ref.~\onlinecite{note3}, specifies
the neighbor  class.
Note that $n{=}1$ and $n{=}2$ are two inequivalent  classes of NN's.
As can be seen in Fig.~\ref{f:wurtzite},  $n{=}1$ corresponds to ``in plane''
NN's (i.e., NN's which are in the same $c$-plane), whereas  $n{=}2$
corresponds to ``out of plane'' NN's.
The distances in Table~\ref{table1} are for the ideal hcp structure,
with $c/a{=}\sqrt{8/3}$, where $a$ is the NN distance.
A neighbor of the symmetry  class $n{=}3$ corresponds to a second
neighbor in the classification by distance.
Neighbors of class $n {=} 4$ are reached from the central cation by moving a
distance $c$ along the $c$-axis.
They are the closest neighbors along the hexagonal direction.
Neighbors  of  classes $n{=}5$ and $n{=}6$ are equidistant but are
inequivalent by symmetry.
The remaining neighbor classes in Table~\ref{table1}, $n{=}7$ and $n{=}8$,
are included in Fig.~\ref{f:wurtzite}.
The parameter $z_n$ in Table~\ref{table1} is the ``coordination number,''
i.e., the number of neighbors of  class $n$ surrounding the central cation.

The notation for the exchange constants $J$'s associated with different
neighbors has been evolving, to accommodate newer classifications of
these $J$'s.
In early works, when neighbors were classified by their distance $r$, the
notation for the $J$'s also was based on distance: $J_1$ for NN's, $J_2$ for
next-nearest(second)neighbors, $J_3$, for third neighbors, etc.
For II-VI DMS's with the zinc-blende structure (fcc cation lattice) this
notation is still quite useful because each of the eight shortest distances
$r$ is associated with a unique neighbor class $n$
(see footnote 115 in Ref.~\onlinecite{Shapira02jap}).
However, the distance-based notation is totally inadequate for DMS's with the
wurtzite structure (hcp cation structure).
In the ideal hcp structure, the shortest distance $r$ already corresponds to
two classes of NN's, with different $J$'s.

An early apparent advantage of the distance-based notation  followed from the
prediction\cite{Larson88} that the magnitudes (sizes) of the $J$'s decrease
monotonically with increasing $r$.
If true, this prediction would have made the classification by distance
equivalent to a classification by size.
However, later theories,\cite{Yu95,Wei93} and recent
experiments,\cite{Bindi98prl,Hennion02} indicate that there is no simple
correspondence between size and distance.

The different notations for the $J$'s that are used in the present work serve
different needs.
The simplest notation, $J(n)$, associates the $J$'s with the neighbor classes
$n$ listed in Table~\ref{table1}.
For example, $J(4)$ is the exchange constant with a neighbor of the class
$n{=}4$.
The disadvantages of this notation is that neither the relevant distance $r$
nor the ranking by size are immediately obvious.

An alternative notation, similar to that in Ref.~\onlinecite{Larson86}, is
given in the fourth row of Table~\ref{table1}.
This notation too is based on  division of neighbors into symmetry classes,
but it also gives some information about  the distance $r$.
Instead of the number $n$, the neighbor  class is specified by a combination
of subscripts and superscripts.
The information about distance is given by the numerical value of the
subscript, which increases with increasing $r$.
For the same subscript, the superscripts ``in'' and ``out'' are used to
distinguish between equidistant but inequivalent neighbor classes.
Thus, $J_1^\mathrm{in}$ and $J_1^\mathrm{out}$ are the $J$'s for the two
classes of NN's.
A prime is added as a superscript to indicate that the distance $r$ is
approximately, but not exactly, the same as for an unprimed exchange constant
with the same subscript.
For example, the exchange constants for the neighbor  classes $n{=}7$ and
$n{=}8$, whose distances differ by only 4\%, are designated as $J_4^\prime$
and $J_4$, respectively.

\subsection{Classification of Exchange Constants by Size}
The preceding two notations for the $J$'s were both based on the division of
neighbors into symmetry \emph{classes}.
This classification, however, has a serious practical drawback.
Quite often the magnitude (size) of an exchange constant is measured
\emph{before} the neighbor  class with which it is associated is determined.
Prior to such a determination, any notation based on the neighbor class
is not useful.
It is then more practical to adopt a notation that is based primarily on the
ranking of the $J$'s by size.
In the present work, only the four largest $J$'s were measured.
The chosen designations of these $J$'s in terms of their sizes are as follows:
\begin{enumerate}
\item The largest two exchange constants are labeled $J_1$ and $J_1^\prime$,
with $J_1$ chosen (arbitrarily) to be the smaller  of the two.
These two exchange constants are associated with the two inequivalent
classes of NN's.
In the present work it has not been determined which of the two corresponds
to $J_1^\mathrm{in}$ and which to $J_1^\mathrm{out}$.
\item By definition, any exchange constant $J$  which is not associated with
either of the two classes of NN's is a distant-neighbor (DN) exchange
constant.
The largest DN exchange constant (third-largest overall) is called $J^{(2)}$,
the second-largest DN exchange constant (fourth-largest overall)
is called $J^{(3)}$.
\end{enumerate}
The assignment of $J^{(2)}$ and $J^{(3)}$ to specific neighbor classes is a
major task class that will be discussed in detail.

\section{EXPERIMENTAL}
\label{s:exp}
\subsection{Samples}
\label{ss:samples} Single crystals of \ZnMnO\
($0.0056{\le}x{\le}0.030$) were grown by chemical vapor transport
using chlorine as the transporting agent. The growth temperature
was $900^\mathrm{o}\mathrm{C}$. The Mn concentration $x$ was
obtained using three methods:
\begin{enumerate}
\item  From the Curie constant, obtained  from a fit of the
susceptibility between $200$ and $300\;\mathrm{K}$ to a sum of a
Curie-Weiss susceptibility and a constant $\chi_d$ representing
the lattice diamagnetism. The values $S {=} 5/2$  and $g {=}
2.0016$ (Ref.~\onlinecite{Dorain58}) for the $\mathrm{Mn^{2+}}$
ion were used. \item  From the apparent saturation value $M_s$ of
the magnetization. The determination of $M_s$ (also known as the
``technical saturation value'') is discussed in Sec.~\ref{s:vsm}.
The relation between $x$ and $M_s$ was discussed in
Refs.~\onlinecite{Shapira02jap} and \onlinecite{Shapira90jap},
among others. This relation is based on the assumption of a random
distribution of the Mn ions over the cation sites. \item From
atomic emission spectroscopy with inductively coupled plasma
(ICP-AES).
\end{enumerate}

Table~\ref{table2} compares the results of the three methods. The
good agreement indicates that the apparent saturation value $M_s$
is consistent with a random distribution of the Mn ions. The last
column in Table~\ref{table2} gives the chosen values of $x$ that
will be used to label the various samples, namely, $x {=} 0.0056,
0.021, 0.029, \textrm{\ and\ } 0.030$. These values will also be
used in the data analysis.

X-ray powder diffraction data were obtained on small portions of
the samples with $x {=} 0.0056, 0.021 \textrm{\ and\ } 0.030$.
These data were taken with a Bruker model ``D8 Discover with
GADDS'' spectrometer, using Cu-$\mathrm{K}_\alpha$ radiation. All
the powder diffraction patterns were in good agreement with the
wurtzite structure (space group $P6_3mc$). No other
crystallographic phase was detected.

The samples used in measurements of the magnetization $M$, and of
the differential susceptibility $dM/dH$, had linear dimensions of
$2$ to $4\;\mathrm{mm}$. The only exception was one set of
pulsed-field data on a powder obtained by crushing the single
crystal with $x{=} 0.029$. Pulsed-field data on that single
crystal were obtained before it was crushed.

\begin{table}
  \caption{\label{table2}
  Mn concentration, $x$, as obtained from:
  1) the apparent saturation value $M_s$,
  2) the susceptibility between $200$ and $300\;\mathrm{K}$, and
  3) atomic emission spectroscopy with inductively coupled plasma
  (ICP-AES).
  The ``best value'' (B.V.) is the value adopted in the text.
}
\begin{ruledtabular}
  \begin{tabular}{cdddd}
Sample &  \multicolumn{1}{r}{$x$($M_s$)}
       &\multicolumn{1}{r}{$x$(Suscept.)}
       &\multicolumn{1}{r}{$x$(ICP-AES)}
       &\multicolumn{1}{r}{$x$(B.V.)}\\
       \hline
A   &0.0056  &0.0057  &0.005   &0.0056\\
B   &0.0210  &0.0208  &0.021   &0.021\\
C   &0.0291  &0.0286  &\multicolumn{1}{c}{---}      &0.029\\
D   &0.0305  &0.0283  &0.030   &0.030\\
  \end{tabular}
\end{ruledtabular}
\end{table}

\subsection{Magnetization measurements}
\label{ss:mag}
Three  types of magnetization measurements were performed:
\begin{enumerate}
\item The magnetization $M$ at $T {=} 20\;\mathrm{mK}$ was measured in dc
magnetic fields up to $90\;\mathrm{kOe}$.
These data were taken with a force magnetometer operating in a plastic
dilution refrigerator.
The experimental techniques were described earlier.\cite{Bindi96cjp}
The magnetic field $\mathbf{H}$ was either parallel to the $c$-axis,
or parallel to the $[10\bar{1}0]$ direction (one of the directions
perpendicular to the $c$-axis.)
The dc magnetic field was produced by a NbTi superconducting magnet.
Several traces of $M{\textit{\ vs.\ }}H$, in both increasing and
decreasing $H$, were taken for each experimental configuration.
All such traces were similar, and showed no hysteresis.
They were averaged in order to improve the signal/noise ratio in the
final result.
\item The magnetization $M$ was measured at $0.65\;\mathrm{K}$ in dc
fields up to $170\;\mathrm{kOe}$.
The samples were immersed in a liquid $\mathrm{{}^3He}$ bath.
A vibrating sample magnetometer (VSM) operating in an 18-tesla superconducting
magnet ($\mathrm{Nb_3Sn}$ wire) was used.\cite{Gratens01prb}
Again, no hysteresis was observed, and traces were averaged to improve the
signal/noise ratio.
\item The differential susceptibility $dM/dH$ was measured in pulsed
magnetic fields up to $500\;\mathrm{kOe}$.
The experimental techniques were described earlier.\cite{Foner89}
The pulse duration  was $7.4\;\mathrm{ms}$.
The sample was in direct contact with a liquid $\mathrm{{}^4He}$ bath
maintained at a temperature $T_\mathrm{bath} {=} 1.5\;\mathrm{K}$.
However, the data showed that despite the direct contact, the sample was not
in thermal equilibrium with the liquid-helium bath during the pulse.
Such non-equilibrium effects were found in earlier pulsed-field
experiments.\cite{Foner94,Shapira99,Shapira01,Ajiro98}
\end{enumerate}

\section{NN EXCHANGE CONSTANTS FROM PULSED-FIELD DATA}
\label{s:pulse}
\subsection{NN Exchange Constants}
\label{ss:nn}

\begin{figure}
\includegraphics[scale=0.4, keepaspectratio=true, clip]{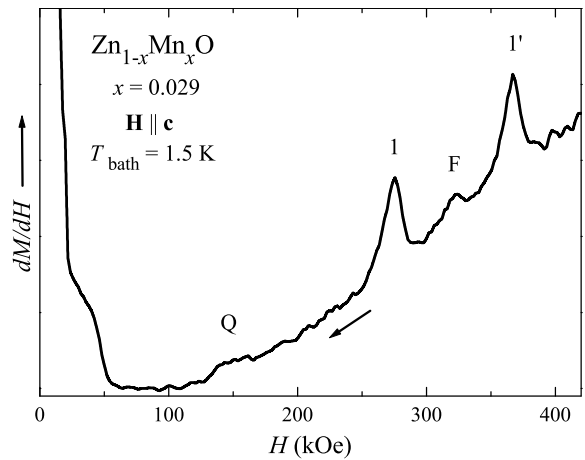}\\
\caption{\label{f:pulse} Differential susceptibility $dM/dH$ for a
single crystal with $x {=} 0.029$, measured in pulsed fields.
These results are for the ``down'' portion of the pulse
(decreasing $H$), with the magnetic field parallel to the
$c$-axis. The two large peaks, $1$ and $1^\prime$, are attributed
to the two inequivalent NN pairs. The  small peaks Q and F are
discussed in the text.}
\end{figure}

\begin{figure}
\includegraphics[scale=0.4, keepaspectratio=true, clip]{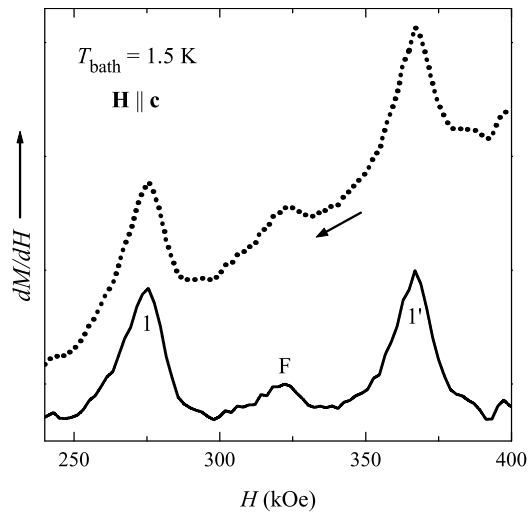}\\
\caption{\label{f:phigh}
The high field portion of the results in Fig.~\ref{f:pulse}.
The raw experimental data, from Fig.~\ref{f:pulse}, are represented by the
dotted curve.
The solid curve was obtained by subtracting a linear baseline.
The two dominant peaks $1$ and $1^\prime$ are attributed to the two
inequivalent NN pairs in the wurtzite structure.}
\end{figure}

Figure~\ref{f:pulse} shows $dM/dH {\textit{\ vs.\ }} H$ for the
single crystal with $x {=} 0.029$. This trace is from the
field-down portion of a pulse with a maximum field of
$500\;\mathrm{kOe}$. The two prominent peaks at high fields,
labeled as $1$ and $1^\prime$, correspond to two MST's. The
expanded view of these two peaks, shown in Fig.~\ref{f:phigh},
indicates that both peaks have similar heights and widths. The
maxima are at $H_1{=}275\;\mathrm{kOe}$ and
$H_1^\prime{=}367\;\mathrm{kOe}$. These two  peaks were not
resolved in the up portion of this field pulse.

Pulsed-field measurements were also performed on single crystals
with $x {=} 0.021$ and $0.030$, and on a powder obtained by
crushing the single crystal  with  $x{=}0.029$ (after the data in
Figs.~\ref{f:pulse} and \ref{f:phigh} were obtained). In all cases
the large peaks, $1$ and $1^\prime$, were resolved in the down
portion of the pulse. However, in the up portion of the pulse
these peaks were well resolved only in the following situations:
a) for the powder sample, in all field pulses, and b) for the
single crystal with $x{=}0.029$ when the maximum field was
$420\;\mathrm{kOe}$, as compared to $500\;\mathrm{kOe}$ for the
pulse in Fig.~\ref{f:pulse}. These results are explained later.

Based on the data in all samples, the two large MST's observed in pulsed
fields are at $H_1{=}270{\pm}8\;\mathrm{kOe}$ and
$H_1^\prime{=}362{\pm}8\;\mathrm{kOe}$.
Because the uncertainties in these two values are correlated, the uncertainty
in the difference is much smaller, i.e.,
$(H_1^\prime{-}H_1){=}92{\pm}2\;\mathrm{kOe}$.
The average is $\overline{H}_1{=}(H_1^\prime+H_1)/2{=}316{\pm}8\;\mathrm{kOe}$.

Earlier data for other DMS's with the wurtzite
structure\cite{Shapira89,Bindi92} showed that: a) each MST from NN
pairs splits into a doublet, corresponding to the two inequivalent
classes of NN's, and b) the two corresponding exchange constants,
$J_1^\prime$ and  $J_1$, are the largest. The two large MST's at
$H_1^\prime$ and $H_1$ are therefore attributed to the two
inequivalent NN pairs.\cite{note27}

For $\mathrm{Mn^{2+}}$ pairs with intra-pair exchange constant $J$, the
magnetic fields $H_n$ at the  MST's are given by\cite{Shapira02jap}
\begin{equation}
g\muB H_n =2n|J|,                   \label{eq1}
\end{equation}
where $n {=} 1,2,\dots,5$.
In the present case the calculated deviations from Eq.~(\ref{eq1}), caused by
anisotropies and DN interactions, turn out to be smaller than the experimental
uncertainties in $H_1$ and $H_1^\prime$.
Using $n{=}1$ and $g{=}2.0016$, we obtained
$J_1/\kB{=}{-}18.2{\pm}0.5\;\mathrm{K}$ and
$J_1^\prime/\kB{=}{-}24.3{\pm}0.5\;\mathrm{K}$.
Unfortunately, it was not possible to conclude which of the two exchange
constants is $J_1^\mathrm{in}$ and which is $J_1^\mathrm{out}$.
In the case of \dms{Cd}{Mn}{Se},\cite{Bindi92} the smaller of the two NN
exchange constants, defined here as $J_1$, was identified as
$J_1^\mathrm{out}$, and the larger as $J_1^\mathrm{in}$.
This identification was based on the effect of the Dzyaloshinkii-Moriya (DM)
interaction on the widths of the MST's, and it also agreed with Larson's
prediction.\cite{Larson90}
In the present work the effect of the DM interaction was not apparent in
the data, presumably because this interaction decreases rapidly as the atomic
number of the \emph{anion} decreases, i.e., much smaller for oxygen than for
selenium.\cite{Larson89}
Although direct evidence is lacking, based on the experimental result for
\dms{Cd}{Mn}{Se}\ and on the theory, we speculate that in the present material
too, $J_1{=}J_1^\mathrm{out}$ and  $J_1^\prime{=}J_1^\mathrm{in}$.

As discussed earlier,\cite{Bindi92} the dominant superexchange path for both
classes of NN's, which is through the intervening anion, is the same.
The difference between $J_1^\mathrm{in}$  and   $J_1^\mathrm{out}$ is
attributed to differences in the other exchange paths whose contribution
is smaller.
In the present material
$\Delta\overline{J}_1{=}|J_1^\mathrm{in}{-}J_1^\mathrm{out}|$ is $29\%$ of the
average $\overline{J}_1/\kB {=} {-}21.2{\pm}0.5\;\mathrm{K}$.
This percentage difference should be compared to $13\%$ in \dms{Cd}{Mn}{S},
and $15\%$ in \dms{Cd}{Mn}{Se}.\cite{Shapira89,Bindi92}

\subsection{Other Features of the Pulsed-Field Data}
\label{ss:other} Figure~\ref{f:pulse} also shows two small peaks,
labeled as Q and F. Computer simulations indicate that these peaks
are due in part to MST's from quartets (tetramers). Each  of these
quartets consists of  four spins that are coupled by some
combination of $J_1$ and/or $J_1^\prime$ exchange bonds. Other
possible contributions to these small peaks may be due to
cross-relaxation processes that can occur in the absence of
equilibrium.\cite{Ajiro98,Wernsdorfer02,Paduan03eprint} The
non-equilibrium behavior in pulsed fields, including
cross-relaxation, will be discussed in a later publication. Very
briefly, the cross relaxation processes that may contribute to Q
are similar to those discussed in Ref.~\onlinecite{Paduan03eprint}
in connection with the second-harmonic peak $\mathrm{P}_{1/2}$.
The cross relaxation process that may contribute to F involves two
NN pairs of different  classes. The latter process is most rapid
at $\overline{H_1}{=}(H_1^\prime+H_1)/2$, where the energy
separation between the two lowest levels for one  class of NN pair
matches that for the other class of NN pair.

The widths of peaks $1$ and $1^\prime$ provide convincing evidence for other
types  of non-equilibrium processes, associated with a phonon-bottleneck,
which restricts the sample-to-bath heat
flow.\cite{Shapira99,Shapira01,Miyashita,Waldmann02,Chiorescu,Inagaki03}
From Fig.~\ref{f:phigh}, the full width at half height of either of these two
peaks is $14\;\mathrm{kOe}$.
This value should be compared with a minimum equilibrium width of
$39\;\mathrm{kOe}$ at $T_\mathrm{bath}{=}1.5\;\mathrm{K}$.
In addition to narrow widths, peaks $1$ and $1^\prime$ also show a pronounced
asymmetry which is characteristic of non-equilibrium behavior resulting from
a phonon bottleneck.
The $14\kOe$ width at half height is the sum of $5.5\kOe$ from the rise and
$8.5\kOe$ from the fall.
(The data in Fig.~\ref{f:phigh} are for decreasing $H$, so that the rise
corresponds to fields above the maximum of $dM/dH$, and the fall corresponds
to fields below the maximum.)
Both the narrowing and the asymmetry  are predicted from  models for the
phonon bottleneck.

The phonon bottleneck also accounts for the difficulty of resolving the peaks
$1$ and $1^\prime$ during the up portion of the field pulse.
At the beginning of the pulse the magnetocaloric effect associated with the
alignment of the singles causes the sample to warm.
This heating is basically the inverse of cooling by adiabatic demagnetization,
except that it is not fully adiabatic.
To observe the peaks $1$ and $1^\prime$, much of this heat must be transferred
to the bath before these peaks are reached.
In the up portion of the pulse the peaks are reached earlier than in the
down portion.
Apparently, in the case of the single crystals an insufficient amount of heat
was transferred before reaching these peaks on the way up.
The peaks were therefore not resolved in the up portion of the pulse.
Lowering the maximum field of the pulse leads to a slight delay of the time
when the peaks are reached on the way up, which improves the chance of
resolving these peaks.
A more drastic change occurs when a single crystal is crushed into powder.
The resulting much larger surface/volume ratio improves the sample-to-bath
heat flow substantially, and the peaks are also resolved in the up portion of
the pulse.

Another phenomenon seen in Fig.~\ref{f:pulse} is that (for decreasing $H$) a
rapid rise of $dM/dH$ occurs below about $50\kOe$.
This rise is due to MST's from clusters in which the spins are coupled by DN
exchange constants.
The determination of these DN exchange constants is the main topic in the
remainder of the paper.

\section{DC MAGNETIZATION AT 0.65 K}
\label{s:vsm}
\begin{figure}
\includegraphics[scale=0.4, keepaspectratio=true, clip]{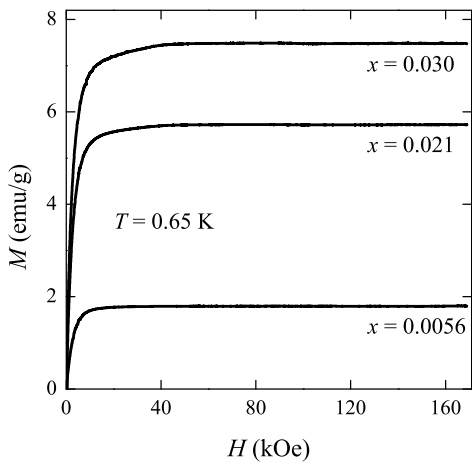}\\
\caption{\label{f:vsm} Magnetization traces for \ZnMnO\ crystals
with $x{=}0.0056$, $0.021$ and $0.030$ at $T{=}0.65\K$. These data
were taken in dc magnetic fields. A (minor) correction for the
lattice diamagnetism is included.}
\end{figure}

Figure~\ref{f:vsm} shows  magnetization curves at $0.65\K$,
measured in dc magnetic fields up to $170\kOe$. As already noted,
there was no hysteresis in any of the data taken in dc fields. The
curves in Fig.~\ref{f:vsm} exhibit the expected
behavior.\cite{Shapira02jap} Above $50\kOe$ the magnetization $M$
shows an apparent saturation (``technical saturation''). The
apparent saturation value $M_s$ is lower than the true saturation
value $M_0$. The latter is expected to be reached in fields
substantially above $170\kOe$. The relation between $M_s$ and $x$
was used in Sec.~\ref{s:exp} as one of the three methods of
determining $x$. An expanded view of the upper portion of each of
the curves in Fig.~\ref{f:vsm} does not show any MST between $50$
and $170\kOe$. (The magnetization change associated with the small
peak Q in Fig.~\ref{f:pulse} is estimated to be about $0.2\%$.
This small change was not resolved in any of the dc data for
$x\leq0.03$). The absence of detectable MST's in the dc data
between $50$ and $170\kOe$ indicates that all MST's from DN pairs
occur below $50\kOe$.

A feature of Fig.~\ref{f:vsm} which is most obvious for
$x{=}0.030$ is a magnetization ``ramp'' ending slightly above
$40\kOe$. Magnetization ramps are produced by the coalescence of
broadened MST's.\cite{Shapira02jap} The ramp ending just above
$40\kOe$ is due to the coalescence of MST's from clusters
involving the largest DN exchange constant, defined earlier as
$J^{(2)}$. These MST's  were not resolved at $0.65\K$, but were
resolved at $20\mK$.

\section{DN EXCHANGE CONSTANTS AND SINGLE-ION ANISOTROPY FROM 0.02 K DATA}
\label{s:fridge}
\subsection{Overall View of the dc Magnetization at 20~mK}
\label{ss:overall}
\begin{figure}
\includegraphics[scale=0.4, keepaspectratio=true, clip]{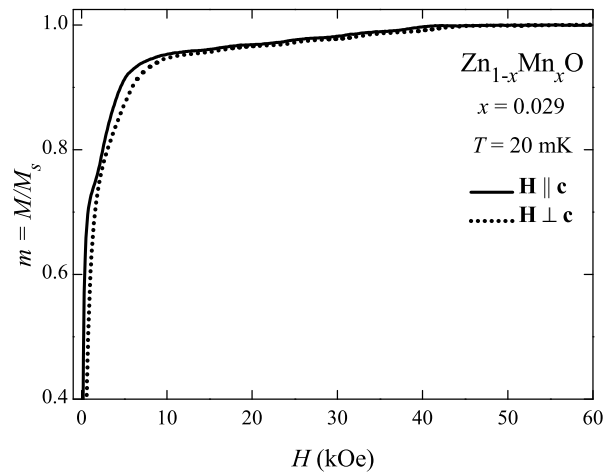}\\
\caption{\label{f:dilm29} Magnetization traces for $x{=}0.029$
measured at $T{=}20\mK$ with $\mathbf{H}\perp\mathbf{c}$ (along
the $[10\bar{1}0]$ direction) and $\mathbf{H}\parallel\mathbf{c}$.
The magnetization was corrected for the lattice diamagnetism and
normalized to its technical saturation value $M_s$. }
\end{figure}

\begin{figure}
\includegraphics[scale=0.4, keepaspectratio=true, clip]{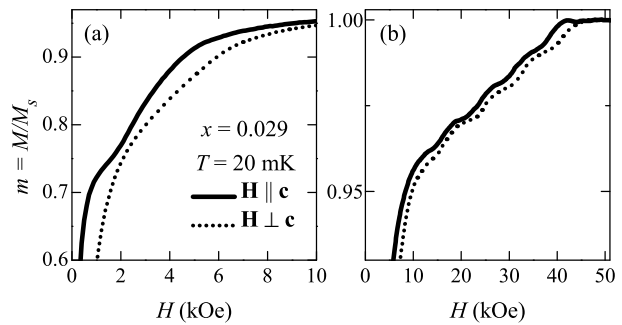}\\
\caption{\label{f:dilm29detail}
Expanded views of (a) the low field portion, and (b) the high field portion
of the magnetization traces in Fig.~\ref{f:dilm29}.
}
\end{figure}

Figure~\ref{f:dilm29} shows $20\mK$ data for $x{=}0.029$, taken
both with $\mathbf{H}\parallel\mathbf{c}$  and
$\mathbf{H}\perp\mathbf{c}$. These data are normalized to the
technical saturation value, $M_s$, and are corrected for the
lattice diamagnetism. Expanded views of portions of these data are
shown in Fig.~\ref{f:dilm29detail}. The main features are:
\begin{enumerate}
\item $M$ rises quickly at low fields. This initial fast rise is typical.
It is mainly due to the alignment of singles ($\mathrm{Mn^{2+}}$ ions with no
significant exchange coupling to other
$\mathrm{Mn^{2+}}$ ions).\cite{Shapira02jap}
\item The initial fast rise of $M$ is followed by a ramp which ends near $5$
or $6\kOe$, depending on field direction.
This ramp is shown more clearly in the expanded view of
Fig.~\ref{f:dilm29detail}(a).
\item A second ramp, smaller in height but spread over a larger field
interval, follows the first ramp.
The second ramp ends near $40\kOe$.
Just above this field, $M$ reaches technical saturation.
\item The magnetization depends on field direction.
This anisotropic behavior is more obvious at low $H$.
\end{enumerate}

Each of the two ramps is due to the coalescence of  broadened
MST's.\cite{Shapira02jap}
A well defined ramp usually corresponds to one series of MST's.
The high-field end of such a ramp is near the last MST from this series,
and it can be used to estimate the relevant exchange constant $J$.
A more accurate value for  $J$ can be obtained if the MST's on the ramp
are well resolved.

The end of the second ramp, near $40\kOe$, leads to the estimate
$J^{(2)}/\kB{\approx}{-}0.5\K$ for the largest DN exchange constant.
The end of the first ramp, near $6\kOe$, gives
$J^{(3)}/\kB{\approx}{-}0.08\K$.
To improve on these rough estimates it is necessary to examine the MST's
which give rise to each ramp, taking into account the relevant weak
anisotropic interactions.

\subsection{Cluster Models and Anisotropies}
\label{ss:models}
Cluster models play a key role in analysis of MST's involving DN exchange
interactions.\cite{Shapira02jap}
Detailed information about cluster models and their statistical properties is
given in Ref.~\onlinecite{note3}.
In the present work, only the largest two DN exchange constants, $J^{(2)}$
and $J^{(3)}$, were determined.
The cluster models that were used in the data analysis depended on the field
range of the data:
\begin{enumerate}
\item  For fields below $8\kOe$, both $J^{(2)}$ and $J^{(3)}$ are important.
These exchange constants correspond to two  classes of DN's, but the identity
of neither of these two  classes was known at the beginning of the analysis.
Therefore, the cluster models used for this field range included:
a) the two  classes of NN's, and b) any possible two  classes of DN's selected
from the six classes listed in Table~\ref{table1}.
\item Because the magnetization above $10\kOe$ is hardly affected by
$J^{(3)}$, the cluster models for the field range from $10$ to $50\kOe$
involved only the one  class of a DN associated with $J^{(2)}$, and the two
NN classes associated with $J_1$ and $J_1^\prime$.
All six possible choices for the class for the  DN were tried.
\item Analysis of the data between  $50\kOe$ and $90\kOe$ was based primarily
on a model which included only the two NN exchange constants.
Simulations of the magnetization curves showed that in this field range the
effects of the DN exchange constants $J^{(2)}$  and  $J^{(3)}$ were very small.
\end{enumerate}

In addition to the exchange interactions, two anisotropies were also included
in the analysis: single-ion anisotropy, and dipole-dipole (dd) anisotropy.
The effects of these anisotropies on the MST's (or ramp) from pairs become
more pronounced as the magnitude $|J|$ of the relevant intra-pair exchange
constant decreases.
That is, the effects caused by the anisotropies are very small for the MST's
originating from NN pairs of either  class;
small but easily detected for the MST's  (ramp) from pairs involving \J{2};
and  very pronounced for the MST's (ramp) from pairs involving $J^{(3)}$.

The single-ion anisotropy can be described by the hamiltonian
\[DS_z^2 + \frac{a}{6}(S_x^4+S_y^4+S_z^4),\]
where  the $z$ direction is along the $c$-axis.
From Electronic Paramagnetic Resonance (EPR) at $77\K$, $D/\kB{=}{-}31\mK$ and
$a/\kB{=}0.3\mK$.\cite{Dorain58}
Because the term involving $a$ is relatively small, it was neglected in the
present work.
Values  of the dd anisotropy constant for different  classes of pairs are
given in Table~\ref{table1}.
They are based on a simple model in which the two spins in a pair are
represented by two points separated by $r_n$.

\subsection{Results at 20 mK and their Analysis}
\subsubsection{Objectives}
The two objectives of the experiments at $20\mK$ were:
1) to determine the values of  $J^{(2)}$, $J^{(3)}$, and $D$,   and
2) to identify the DN  classes associated with $J^{(2)}$ and $J^{(3)}$.
The analysis consisted of a number of steps taken in  sequence.
In what follows, each step in this sequence is outlined, and the results
are summarized.
The main assumption in the analysis is that the  Mn ions are randomly
distributed over the cation sites.

\subsubsection{Procedures for Identifying DN  classes}
The general procedure of  associating DN constants of known magnitudes with
different possible classes of DN's involves comparisons between experimental
magnetization curves and computer simulations.\cite{Shapira02jap,Bindi98prl}
Separate simulations are  carried out for all competing possibilities for the
DN  classes.
The six DN  classes listed in Table~\ref{table1} lead to $6{\times}5{=}30$
possibilities for the two DN  classes associated with \J{2}\ and \J{3}.
(Interchanging the order of the two  classes of DN's leads to a new
possibility.)

In the present work these laborious simulations were postponed,
because a preliminary identification of the DN  classes was
possible based on a simpler, and more physical, procedure. Two
favorable circumstances permitted the simpler procedure: a large
difference in the magnitudes of \J{2}\ and \J{3}, and the
availability of data for a sample with a very small $x$.

For the lowest Mn concentration, $x{=}0.0056$, the ramp which ends
near $40\kOe$ is due primarily to \J{2}\ pairs. To a good
approximation, the total magnetization rise $\Delta M^{(2)}$
associated with this ramp is therefore related to the fraction of
Mn ions which are in such pairs. This fraction can be calculated
for any possible choice of the DN  class associated with \J{2}, so
that $\Delta M^{(2)}$ for any choice can be calculated and
compared with experiment. Inclusion of \J{2}\ triplets, in
addition to \J{2}\ pairs, in the calculation of $\Delta M^{(2)}$
makes the comparison even more reliable.

Using the statistical tables in Ref.~\onlinecite{note3}, the
contribution of pairs and triplets to $\Delta M^{(2)}$ was
calculated for each of the six possible choices of the DN
associated with \J{2}. Only one of these choices  agreed with the
experimental value of $\Delta M^{(2)}$. Thus, a unique
identification of the cluster  class associated with \J{2} was
achieved. A similar procedure was also used for a preliminary
identification of the DN responsible for \J{3}, but in that case
two possible DN  classes gave good agreement with experiment,  so
that a unique choice  could not be made. The full numerical
simulations, \emph{for all the samples}, carried out at the end of
the analysis, confirmed the preliminary identifications of the DN
classes on the basis of the data for $x{=}0.0056$.

\subsubsection{DN  Class Responsible for \J{2}}
\begin{figure}
\includegraphics[scale=0.4, keepaspectratio=true, clip]{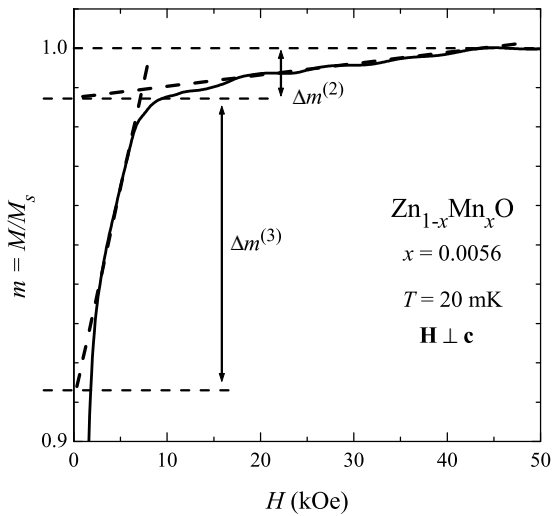}\\
\caption{\label{f:deltaM} Magnetization trace for $x{=}0.0056$,
obtained at $T{=}20\mK$ with $\mathbf{H}\perp\mathbf{c}$. The
ordinate $m$ is the magnetization $M$ normalized to the technical
saturation value $M_s$. This figure illustrates the procedure used
to estimate the contribution $\Delta m^{(2)}$ to $m$ from all
\J{2} clusters, and the contribution $\Delta m^{(3)}$ to $m$ from
all \J{3} clusters.}
\end{figure}

The procedure of extracting the experimental ratio $\Delta
m{=}\Delta M/M_s$ from the data, for the ramps associated with
\J{2}\ and \J{3}, is illustrated in Fig.~\ref{f:deltaM}. For the
higher-field ramp, associated with \J{2}, it gives $\Delta
M^{(2)}/M_s {=} 1.3\%$ for $x{=}0.0056$. The theoretical value of
$\Delta M^{(2)}/M_s$ is obtained by multiplying $(\Delta
M^{(2)}/M_0)$ by $(M_0/M_s)$. The first ratio depends on the DN
class $n$ that corresponds to \J{2}. For each possible $n$, this
ratio  was obtained from  probability tables for pairs and
triplets that involve only the exchange constant for class $n$
[\onlinecite{note3}]. The probabilities are   based on a cluster
model that includes only the two classes of NN's and the DN class
$n$. The second ratio, $(M_0/M_s)$, was calculated  from the
so-called NN cluster model,\cite{Shapira02jap} in which  only the
two  classes of NN's are included.

For small $x$ the calculated ratio of  $\Delta M^{(2)}/M_s$
increases as the coordination number $z_n$  increases. This
dependence on $z_n$ is the key for identifying the DN  class  that
corresponds to \J{2}. For $x{=}0.0056$ the calculated values of
$\Delta M^{(2)}/M_s$ are: $1.39\%$ for the only DN  class with
$z_n{=}2$; between $3.1\%$ and $4.2\%$ for the three possible
classes with $z_n{=}6$; and $6.1\%$ for the two possible  classes
with $z_n{=}12$. The experimental value $1.3\%$ agrees only with
$z_n{=}2$, which corresponds to $n{=}4$ in Table~\ref{table1}. In
the ideal wurtzite structure  such  a  DN is reached from the
``central'' cation by moving a distance $r_4 {=} c {=}\sqrt{8/3}a$
along the $c$-axis (see Fig.~\ref{f:wurtzite}). In
Table~\ref{table1} the exchange constant $J(4)$ for $n{=}4$ is
designated as $J_3^\prime$. It is noteworthy that this largest DN
exchange constant is not  for the DN that is closest to the
central cation. The closest DN is of  class $n{=}3$, at a distance
$r_3{=}\sqrt{2}a$. Thus, the magnitudes of the exchange constants
do not decrease monotonically with distance. This non-monotonic
dependence on distance was  predicted by some
theories,\cite{Yu95,Wei93} and has been observed experimentally
earlier.\cite{Bindi98prl,Hennion02}

\subsubsection{\label{ss:values}Values of  $J_3^\prime$ and $D$}
\begin{figure}
\includegraphics[scale=0.4, keepaspectratio=true, clip]{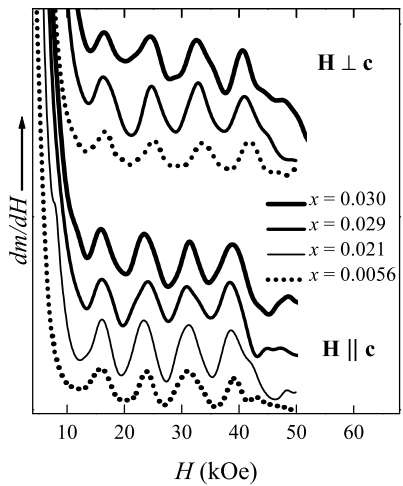}
\caption{\label{f:dmdH} Numerical derivatives $dm/dH$ of $m\equiv
M/M_s$, obtained from the experimental magnetization traces at
$20\mK$. The results are for two field directions:
$\mathbf{H}\parallel\mathbf{c}$ and
$\mathbf{H}\parallel[10\bar{1}0]$. The latter direction is
designated as $\mathbf{H}\perp\mathbf{c}$. The sample with
$x=0.021$ has been measured only with
$\mathbf{H}\parallel\mathbf{c}$. The traces have been displaced
vertically relative to each other, but the gain is the same. }
\end{figure}

The values for $J_3^\prime$ and $D$ were obtained from analysis of the well
resolved MST's on the ramp associated with $J^{(2)}{=}J_3^\prime$
(see part (b) of Fig.~\ref{f:dilm29detail}).
These MST's stand out more clearly in the derivative $dm/dH$ of the normalized
magnetization $m\equiv M/M_s$, shown in Fig.~\ref{f:dmdH}.
For three of the four samples in this figure the data are for both
$\mathbf{H}\perp\mathbf{c}$ and $\mathbf{H}\parallel\mathbf{c}$.
The dependence of the fields at the MST's on field direction is caused by the
anisotropy.
Because the $c$-axis is an easy axis ($D{<}0$), the spins in the pairs are
aligned more quickly  for $\mathbf{H}\parallel\mathbf{c}$ than for
$\mathbf{H}\perp\mathbf{c}$.
The faster alignment for $\mathbf{H}\parallel\mathbf{c}$ is more apparent
in part (b) of Fig.~\ref{f:dilm29detail}.

All curves in Fig.~\ref{f:dmdH} show the last four MST's from
$J_3^\prime$ pairs. For $x{=}0.0021$ the first MST is also seen,
but only as a ``shoulder'' on the fast drop of $dm/dH$ at low
fields. The fields at the last four MST's are all above $15\kOe$.
Values of both $J_3^\prime$ and $D$ were obtained from an analysis
of these fields, based on a pair hamiltonian which included the
exchange interaction due to $J_3^\prime$, the uniaxial anisotropy
governed by $D$, and the dd-interaction. The latter was calculated
using the value in Table~\ref{table1} for $n{=}4$.

In the first step of the analysis, an approximate value for
$J_3^\prime$ was obtained from the field at the last (fifth) MST, and the EPR
value of $D$ (Ref.~\onlinecite{Dorain58}) was adopted as the initial value.
Later, the predicted fields at the last four MST's, for both field directions,
were calculated for many sets of ($J_3^\prime$, $D$).
The best match with the experimental values gave:
$J_3^\prime/\kB{=}{-}0.543{\pm}0.005\K$, and $D/\kB{=}{-}0.039{\pm}0.008\K$.
The presumably more accurate  EPR value for $D/\kB$ is ${-}0.031\K$.
The difference may be related to our use of the dd-interaction constant
given in Table~\ref{table1}.
As noted, this constant was obtained from a simple model in which the spins
in a pair are represented by two points separated by $r_4$.

\subsubsection{Possible DN  Classes Associated with \J{3}}
Possible assignments of  the DN  class which corresponds to \J{3}\
were made on the basis of the magnitude $\Delta m^{(3)}$ in
Fig.~\ref{f:deltaM}. The experimental results give $\Delta
m^{(3)}=\Delta M^{(3)}/\Delta M_s{\approx}7.3\%$ for $x{=}0.0056$.
This $\Delta m^{(3)}$ was attributed to the combined magnetization
rise from pairs and triplets involving this exchange constant. The
inclusion of triplets in the theoretical calculation of $\Delta
m^{(3)}$ was more important than in the calculation of $\Delta
m^{(2)}$, because the triplets/pairs population ratio was higher.
The triplets/pairs ratio increases with $z_n$. The lowest possible
coordination number, $z_n{=}2$, is for the neighbor class $n{=}4$
associated with \J{2} and $\Delta m^{(2)}$.

Theoretical values of $\Delta m^{(3)}$ were obtained from cluster models which
included the three largest exchange constants ($J_1$, $J_1^\prime$,
$J_3^\prime$) and any one of  the DN  classes  with either $z_n{=}6$ or
$z_n{=}12$.
The calculated values are: approximately $3.1\%$ for all three DN classes with
$z_n{=}6$, and approximately $6.0\%$ for the two DN  classes with $z_n{=}12$.
On this basis, $z_n$ is equal to $12$, which  leads  to two possible neighbor
classes: $n{=}6$ or $n{=}7$  (see Table~\ref{table1}).
That is, \J{3}\ is either $J_3^\mathrm{out}$ or $J_4^\prime$.
Another possibility (unlikely, but cannot be ruled out entirely) is that
\J{3}\ is associated with two different   classes, both with $z_n{=}6$, that
just happen to have very nearly equal exchange constants.

\subsubsection{Value of \J{3}}
\begin{figure}
\includegraphics[scale=0.4, keepaspectratio=true, clip]{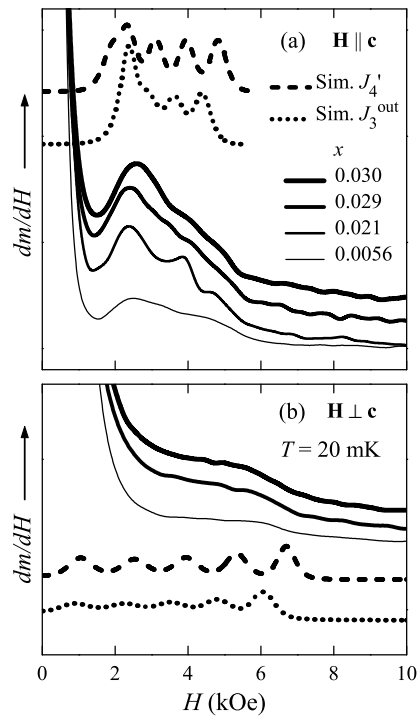}\\
\caption{\label{f:dmdHlowfield} Low field portion of the $dm/dH$ data with
(a) $\mathbf{H}\parallel\mathbf{c}$ and
(b) $\mathbf{H}\perp\mathbf{c}$.
Different  experimental traces are displaced vertically from each other, but
the gain is the same.
Also shown are simulations for the two types of pairs involving the two
classes of neighbors with $z_n{=}12$.
The heavy dashed lines are for pairs with exchange constant $J_4^\prime$
between the central cation and a neighbor of the class $n{=}7$.
The dotted lines are for pairs with $J_3^\mathrm{out}$, involving a
neighbor of the class $n{=}6$.
The simulations are for the actual temperature $T{=}20\mK$, and they
include the single ion anisotropy, involving $D$, and the dd-anisotropy.
 }
\end{figure}

Figure~\ref{f:dmdHlowfield} displays $dm/dH$ data for both
$\mathbf{H}\parallel\mathbf{c}$ and $\mathbf{H}\perp\mathbf{c}$,
in the field range relevant for the analysis of \J{3}. Due to the
smallness of \J{3}, the effects of the anisotropy are important,
as can be judged from the strong dependence of the results on the
direction of $\mathbf{H}$. The $dm/dH$ traces for
$\mathbf{H}\parallel\mathbf{c}$, in Fig.~\ref{f:dmdHlowfield}(a),
do not show the regular sequence of peaks observed in
Fig.~\ref{f:dmdH} for the MST's from $J^\prime_3$-pairs. However,
all traces in Fig.~\ref{f:dmdHlowfield}(a) show a peak slightly
above $2\kOe$. The best resolution is for the sample with
 $x{=}0.021$.  In this sample,
two  additional peaks, just below and just above $4\kOe$ are also
resolved.

The results in Fig.~\ref{f:dmdHlowfield} were compared with
simulations based on a pair hamiltonian which included \J{3}\ and
the two anisotropies. The simulations used $D/\kB{=}{-}39\mK$, as
determined in Sec.~\ref{ss:values}, and the dd-interaction
constant taken from Table~\ref{table1}. Because the latter
constant depends on the DN  class, simulations were carried out
for the two possible neighbor  classes with  $z_n{=}12$, i.e.,
$n{=}6$ with $J_3^\mathrm{out}$,  and $n{=}7$ with $J_4^\prime$.

For $\mathbf{H}\perp\mathbf{c}$, pairs involving neighbors of
either class have different orientations relative to $\mathbf{H}$.
The different orientations lead to different energies from the
dd-anisotropy, and they
 give rise to MST's at slightly different fields.
Therefore, in the simulations for the perpendicular field direction,
 $\mathbf{H}\parallel[10\bar{1}0]$, pairs involving    DN's  of either
of the two possible classes were divided into groups Pairs in
different groups had  different orientations relative to
$\mathbf{H}$. There were three such groups for the DN class
$n{=}6$, with $J_3^\mathrm{out}$, and two groups for $n{=}7$, with
$J_4^\prime$. The simulated magnetization curve  was obtained by
adding the results from all the groups.

The simulations were carried out using the actual temperature,
$T{=}20\mK$. Non-thermal broadening mechanism, such as local
strains that give rise to a spread of the exchange and anisotropy
interactions,\cite{Rubo97} were ignored. In these simulations the
values  $D/\kB {=} -39\mK$, and the dd-interaction constant from
Table~\ref{table1}, were kept fixed. The value of  \J{3}\ was
adjusted to obtain the best match with experiment for both field
orientations. The final $dm/dH$ simulations are shown in
Figs.~\ref{f:dmdHlowfield} (a) and (b) as the dashed and dotted
lines. Obviously, the results of the simulations depend on field
direction. For $\mathbf{H}\parallel\mathbf{c}$, the overall
structure of the experimental $dm/dH$ traces is reproduced by the
simulations. However, because some sources of line broadening were
neglected in the simulations, the detailed structure is better
resolved in the simulations than in the experimental curves. For
$\mathbf{H}\perp\mathbf{c}$ the individual MST's are resolved in
the simulations, even after the different orientations of the
pairs relative to $H$ are included. Experimentally, however, the
individual MST's are not resolved for $\mathbf{H}\perp\mathbf{c}$.
 This difference  is attributed,
again, to the neglect  of some broadening mechanisms in the simulations.
A crude way of accounting
for the neglected broadening mechanisms is to replace the actual
temperature $T$ in the simulations
by a higher effective temperature $T_\mathrm{eff}$.  The minimum
$T_\mathrm{eff}$ which leads to
unresolved MST's for $\mathbf{H}\perp\mathbf{c}$ is $65\mK$.

Some features of the experimental data in
Figs.~\ref{f:dmdHlowfield}(a) and 9(b) are sensitive to the
magnitude of \J{3}. For $\mathbf{H}\parallel\mathbf{c}$ these
features include the field at the most prominent peak, and the
field at the peak associated with the last MST. These peaks stand
out most clearly in the trace for $x{=} 0.021$. For
$\mathbf{H}\perp\mathbf{c}$ the field at the rapid drop of
$dm/dH$, which is at the end of the ramp associated with this
series of MST's, is sensitive to the value of \J{3}. The value of
\J{3} was determined from comparisons of these experimentally
observed features with simulations that used different values of
\J{3}. Assuming that \J{3}\ is $J_3^\mathrm{out}$, the results
gave $J^{(3)}/\kB{=}{-}0.074{\pm}0.005\K$. The alternative,
$J^{(3)}{=}J_4^\prime$  gave ${-}0.082{\pm}0.005\K$. The first choice
gives a slightly better agreement with the data, but in our view
the evidence  is insufficient for concluding that the DN  class is
definitely $J_3^\mathrm{out}$.

\subsubsection{Simulations with both DN exchange constants}
Simulations of the magnetization curves, in fields up to $60\kOe$, were
carried out in order to confirm  the preliminary identifications of the
DN
 classes corresponding to \J{2}\ and \J{3}.
The simulations were for all samples, in contrast with the
preliminary analysis that was carried out only for the sample with
the lowest Mn concentration, $x{=}0.0056$. The simulations used
cluster models which included the two NN exchange constants ($J_1$
and $J_1^\prime$), \J{2}\ and \J{3}, all having the values quoted
above. Each cluster model was based on a specific choice of the
two DN  classes associated with \J{2}\ and \J{3}. As noted
earlier, there are $30$ possible such choices. The results of the
simulations were not sensitive to any change of the DN  class $n$
provided that the coordination number $z_n$ did not change.

Because all anisotropies were neglected in the simulations, the
comparison was
made with ``isotropic'' magnetization curves obtained from the
experimental
data using the relation
\begin{equation}
M_\mathrm{isot}=(M_\parallel+ 2 M_\perp)/3,
\end{equation}
where $M_\parallel$  is for $\mathbf{H}\parallel\mathbf{c}$, and
$M_\perp$  is for $\mathbf{H}\perp\mathbf{c}$.
The widths of the  MST's exhibited by $M_\mathrm{isot}$ are larger than
the
thermal width at $20\mK$ for several reasons:
First,  $M_\mathrm{isot}$ is an average over different groups of pairs
with MST's at slightly different fields.
Second, DN exchange constants smaller than \J{3} were neglected.
Third, variations of the $J$'s, and of the anisotropies, caused by local
strains\cite{Rubo97}
were not included.
To match the widths of the MST's on the ramp which ends at $40\kOe$, the
simulations were carried out using an effective temperature
$T_\mathrm{eff}{=}100\mK$ instead of $20\mK$.
This change has no effect on the identification of the DN's responsible
for
\J{2}\ and \J{3}; the only effect is to smooth the curves.
Simulations using the actual temperature, $20\mK$, lead to the same
identifications of the DN  classes.

\begin{figure}
\includegraphics[scale=0.4, keepaspectratio=true, clip]{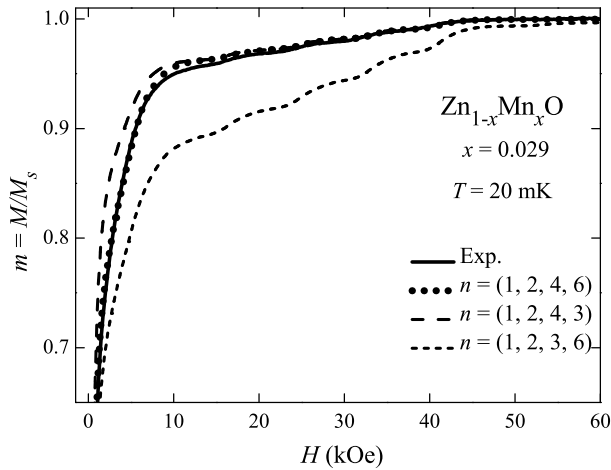}\\
\caption{\label{f:sim}
Comparison between the isotropic ``experimental'' magnetization
$M_\mathrm{isot}{=}(M_\parallel+2M_\perp)/3$, (see text) and numerical
simulations based
on the four exchange constants associated with the neighbor  classes
$n{=}i,j,k,l$.
The simulations assume a random distribution of the Mn ions, and
an effective temperature $T_\mathrm{eff}{=}100\mK$.}
\end{figure}

Figure~\ref{f:sim} shows simulations for $x{=}0.029$. The
``experimental'' curve represents $M_\mathrm{isot}$. Each
simulation is labeled as $n{=}(i,j,k,l)$. This designation means
that the four exchange constants $J_1$, $J_1^\prime$, \J{2},
\J{3}\, that were used in the simulation, correspond,
respectively, to the neighbor  classes $n{=}i,j,k,l$ in
Table~\ref{table1}. All the simulations assumed that the first two
neighbor  classes $(i,j)$ were those of the NN's, that is, $(i,j)$
= $(1,2)$ or $(2,1)$. (These two alternative choices are related
to each other by an interchange of the NN  classes assigned to
$J_1$ and $J_1^\prime$.) Both choices lead to nearly the same
curves  in  this field range. The best agreement is with the
simulation $n{=}(1,2,4,6)$ in which $J^{(2)}{=}J_3^\prime$ and
$J^{(3)}{=}J_3^\mathrm{out}$. This corresponds to one of the two
possibilities which were identified earlier. The other possibility
$n{=}(1,2,4,7)$, not shown in Fig.~\ref{f:sim}, leads to a very
similar curve, so that a definite unique choice of the DN  class
for \J{3}\ is not possible.

The same conclusions concerning the neighbor  classes associated
with \J{2} and \J{3} were reached from comparisons of the data for
the other  samples ($x= 0.0056$, $x=0.021$ and $x=0.030$) with
simulations. Thus, the earlier identifications of the DN  classes
are confirmed. The good agreement between the data and the
simulations also lends support to a random Mn distribution in the
studied samples, which is the main assumption in the simulations.

\begin{acknowledgments}
The work in Brazil was
supported by FAPESP (Funda\c{c}{\~a}o de Amparo {\`a} Pesquisa do Estado
de
S{\~a}o Paulo, Brazil) under contract number 99/10359--7.
The work in Poland was supported by  grant PBZ-KBN-044/P03/2001.
VB and NFOJ acknowledge support from CNPq (Conselho Nacional
de Desenvolvimento Cient{\'\i}fico e Tecnol{\'o}gico, Brazil). Travel
funds
for YS were provided by FAPESP.

\end{acknowledgments}


\begin{thebibliography}{43}
\bibitem[{not({\natexlab{c}})}]{note27}
\bibinfo{note}{Additional evidence that these MST's are from the two dominant
  exchange constants associated with the NN's comes from values of the
  Curie-Weiss temperature $\theta$. These values, obtained both in the present
  study and in Ref.~\onlinecite{Fukumura01}, lead to a dominant AF
  exchange constant which agrees with the more accurate values obtained here
  within $30\%$.}
\end{thebibliography}

\newcommand{\noopsort}[1]{} \newcommand{\printfirst}[2]{#1}
  \newcommand{\singleletter}[1]{#1} \newcommand{\switchargs}[2]{#2#1}

\end{document}